\documentclass{iopart}

\usepackage{amsthm}
\usepackage{latexsym}
\usepackage[dvipdf]{graphicx}
\usepackage{mathrsfs}

\newcommand{\ket}[1]{\left| #1 \right\rangle} 
\newcommand{\bra}[1]{\left\langle #1 \right|} 

\newcommand{\ketbra}[2]{\ket{#1}\!\bra{#2}}

\newcommand{\lie}{{\cal L}}

\newcommand{\tomega}{{\tilde{\omega}}}

\newcommand{\coea}[2]{a^{(#1)}_{#2}}
\newcommand{\tsigma}{{\tilde{\sigma}}}

\begin{document}

\title[Control protocol of finite quantum systems]
{A control protocol of finite dimensional quantum systems using square pules}

\author{Jianju Tang and H. C. Fu \footnote{E-mail: hcfu@szu.edu.cn, corresponding author.}}

\address{School of Physical Sciences and Technology, Shenzhen University,
Shenzhen 518060, P. R. China}

\begin{abstract}
The control protocols of two types of finite dimensional quantum
systems are proposed. The feasibility of each protocol is  possible and an arbitrary target state
can be achieved from initial state by a constant field. The control parameters which are
time periods of interaction between systems and control fields  in each cycles
are connected with the probability amplitudes of target states via trigonometric 
functions and can be determined analytically.
\end{abstract}

\pacs{03.67.Aa, 03.65.Ud, 02.30.Yy, 03.67.Mn}


\section{Introduction}

Quantum control proposed by Huang {\it et.\,al.} in 1983 \cite{huang} is to drive a
quantum system from an initial state to an arbitrary target state through its
interaction with classical control fields or with a quantum accessor. It has
attracted much attention of chemists, physicists and control scientists due to
its application in physics and chemistry, especially in quantum information
processing.  Various notations
in classical control theory were generalized to the quantum control, such as open and
closed control, optimal control \cite{optimal}, controllability \cite{controllability,fu1,fu2,
graph}, feedback control \cite{feedback} and so on. Coherent and incoherent (indirect)
control schemes are proposed. In later case the system is controlled by its interaction
with a quantum accessor which is controlled by classical fields \cite{indirect1,indirect2,
romano,penchen}.
Typically, in the approach of quantum control, one should first model the controlled
system and examine its controllability
which is determined by the system Hamiltonian and interaction Hamiltonian with
classical fields,
and then design classical fields to stream the system to the given target state, which is
referred to as the {\it control protocol} and is the issue we would like to address in
this paper. Some works were proposed along this line, for example, using the Cartan
decomposition of Lie groups \cite{cartan}.

In our previous paper \cite{tang1}, we proposed an explicit control protocol of finite
dimensional quantum system using time-dependent cosine classical field. In this paper, we will
use the square pules to control the finite quantum systems. Advantages of using
square pules is that the interaction Hamiltonian in each control cycle is
time-independent and thus can be treated easily.
Two types of finite systems, the one with equal energy
gaps except the first one and another one with all distinct energy gaps are considered.
The relationship between probability amplitudes of target states and control parameters, the width
of pules, is analytically established.

This paper is organized as follows. In Sec. II, we formulate the controlled system
and control scheme and investigated the controllability. In Sec. III, we present
the control protocol of system with all distinct energy gaps and in Sec. IV we
consider the system with equal energy gaps except one. We conclude in Sec.V.

\section{Control systems and controllability}

\subsection{Control Systems and protocol}

For an $N$-dimensional non-degenerate quantum system with eigen energy $E_n$ and corresponding
eigenstates $|n\rangle$, we suppose that its Hamiltonian described by
\begin{equation}\label{eq2.1}
    H_0=\sum^N_{n=1} E_n \ket{n}\!\bra{n}.
\end{equation}
Without losing generality, we assume $H_0$ is traceless, i.e. $\mbox{tr}H_0 = 0$.
In this paper, just two different types of systems arouse our interest, the first one of which
having all equal energy gaps except the first one, namely
\begin{equation}
\mu_1 \neq \mu_2=\mu_3=\cdots = \mu_{N-1},
\end{equation}
where $\mu_i = E_{i+1}-E_i$ is the energy gap, and the another one with
all district energy gaps
\begin{equation}
\mu_i \neq \mu_j, \qquad i\neq j = 1,2,\cdots, N-1,
\end{equation}
For later convenience, we call them the system I and system II, respectively.
We define energy gaps
\begin{equation}
   \hbar\omega_i = E_{i+1}-E_1, \qquad i=1,2,\cdots,N-1,
\end{equation}
for system I, and
\begin{equation}
   \hbar\tilde{\omega}_i = E_{i+1}-E_i, \qquad i=1,2,\cdots,N-1,
\end{equation}
for system II.

The purpose of this paper is to develop a control scheme to drive the systems to an
arbitrary target state from an initial state, using some independent classical fields
$f_m(t)$. The total Hamiltonian of the system and control fields can be
generally written as
\begin{eqnarray}
H=H_0+H_I,
\qquad
H_I =  \sum_{m=1}^M f_m(t)\hat{H}_m,
\end{eqnarray}
where $M$ is the number of  independent classical fields.

\begin{figure}[t]
  \centerline{\includegraphics[width=10cm]{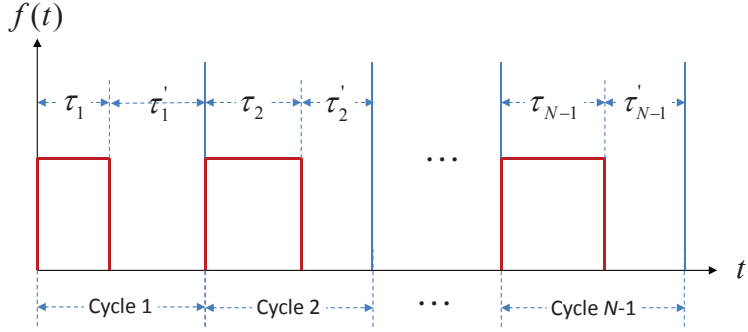}}
  \caption{Control fields.}  \label{fig1-field}
\end{figure}

For the $N$-dimensional systems considered in this paper, the total control
process includes $N-1$ cycles.
In the $m$-th cycle, we first apply a classical field
$
f_m(t)=d_m
$
to control the system for time period $\tau_m$,
and then turn off the control field such that the system evaluates freely for a time
period $\tau_m^\prime$, as
showed in Fig.\ref{fig1-field}.

For system I, It is easy to find that $\hbar\omega_m \neq \hbar\omega_n$ for $m\neq n$.  Thus we let the coupling between the field and system only causing the transition between $E_m$ and $E_1$, which described by the interaction operator
\begin{equation}\label{interaction-opterator-s1}
  H_m=\ketbra{m+1}{1}+\ketbra{1}{m+1}.
\end{equation}

For system II, as  all $\hbar\tomega_m$ are different, we only make the transition between level $E_m$ and $E_{m+1}$
occurs in each cycle. Then the process described by interaction operator
\begin{equation}\label{interaction-opterator-s2}
 H_m=\ketbra{m+1}{m}+\ketbra{m}{m+1}.
\end{equation}
So the control process includes $N-1$ cycles.

For both systems, As the previous discussion, two processes implement in each cycle:
Step (1)  apply the control field to control the system for a time period $\tau_m$ ; Step (2) turn off the control field and allow the system to evaluate for a time period $\tau_m^\prime$. We will
see that the first step provides the real probability amplitude of the target
state and the second step provides the relative phases actually. Therefore, the control field can
be rewritten as
\begin{equation}\label{1-field}
  f_m(t)=\left\{
  \begin{array}{ll}
    d_m , &t_{m-1} \leq t \leq  t_{m-1}+\tau_m \\
     0 , &\mathrm{otherwise},
  \end{array}\right.
\end{equation}
where
$t_m = \sum_{k=1}^{m}(\tau_k +\tau_k^\prime)$.
Those $N-1$ control fields are independent in the sense that each $f_m(t)\neq 0$ in
different time period.

The whole control process can be equivalently regarded as control by one
control field $f(t)=\sum_{m=1}^{N-1} f_m(t)$, where $f(t)$ is shown in
Fig.1. 

\subsection{Complete controllability}

After roughly presentation of the control protocol, we need to examine the controllability of this
control scheme, namely, to examine whether the Lie algebra generated by
the skew-Hermitian operators $iH_0$ and $iH_m$
\begin{equation}
  \lie=\mathrm{Gen}\{ iH_0,iH_m|m=1,2,...,N-1\}
  \label{eq2.4}
\end{equation}
is $\mathrm{su}(N)$ \cite{controllability}. Here $H_m$ for system I and II
are
\begin{eqnarray}
&& H_m = \ketbra{1}{m+1}+\ketbra{m+1}{1}, \\
&& H_m = \ketbra{m}{m+1}+\ketbra{m+1}{m}
\end{eqnarray}
respectively.
Or equivalently, $iH_0, iH_m$ generate the Chevalley basis of su($N$) \cite{fu1,fu2}
\begin{eqnarray}
   && ix_n=i(\ket{n}\!\bra{n+1}+\ket{n+1}\!\bra{n}),\nonumber \\
   && iy_n=\ket{n}\!\bra{n+1}-\ket{n+1}\!\bra{n},\nonumber \\
   && ih_n=i(\ket{n}\!\bra{n}-\ket{n+1}\!\bra{n+1}),
\end{eqnarray}
where $n=1,2,...,N-1$. In fact, it is enough to prove $ix_n\in\lie$ (or $iy_n\in\lie$),
as $iy_n = \mu_n^{-1}[iH_0, ix_n]$ and $ih_n=-[ix_n,iy_n]/2$.

For system I, it is obvious that
\begin{equation}
     ix_1 =iH_1= i(\ketbra{1}{2}+\ketbra{2}{1})\in \lie.
\end{equation}
Then
\begin{eqnarray}
   iy_2&=&[iH_2,iH_1]
        =\ketbra{2}{3}-\ketbra{3}{2}\in\lie, \nonumber  \\
  ix_2&=&\frac{1}{\mu_2}[[iH_2,iH_1],iH_0]=i(\ketbra{2}{3}+\ketbra{3}{2})\in\lie.
\end{eqnarray}
Recursively, we have
\begin{eqnarray}\label{eq2.11}
   iy_{m}&=\mu_m^{-1}\left[[iH_m,iH_{m-1}],iH_0\right].
\end{eqnarray}
So the system I is completely controllable.

For System II, $iH_m$ itself
\begin{eqnarray}
iH_m = i(\ketbra{m}{m+1}+\ketbra{m+1}{m}) ,
\end{eqnarray}
is nothing but the generator $ix_m$.
Therefore, the system II is completely controllable.

\section{Control protocol of system I}
In this section we illustrate the detailed control protocol of System I.
Suppose the system is initially on the ground state $\ket{\psi_0}=\ket{1}$.
The system is driven to an arbitrary target state after $N-1$ cycles.
Relationship between the control parameters $\{\tau_m, \tau_m^\prime\}$
and probability amplitude of target states is explicitly established.

From the protocol we provide, there are two processes implemented in each cycle:
1.fleid on ; 2.field off. In the $m-th$ cycle, We define
\begin{equation}
   \hbar\omega_m=E_{m+1}-E_1
\end{equation}
and let $\frac{1}{2}(E_{m+1}+E_1)=0$ i.e. reset the energy base point.
Then in each cycle, the system can be regard as just having two energy levels.
Using these two conditions,the free Hamilton takes the form:
\begin{eqnarray}\label{h0-hamilton}
  H_0 &= \sum_n^N E_n\ketbra{n}{n}                                                       \nonumber\\
     &=E_1\ketbra{1}{1}+E_ {m+1}\ketbra{m+1}{m+1}
       +\sum^N_ {n=2\atop n\neq m+1}E_n\ketbra{n}{n}                                      \nonumber  \\
     &= -\frac{\hbar\omega_m}{2}\ketbra{1}{1}+\frac{\hbar\omega_m}{2}\ketbra{m+1}{m+1}
        +\sum^N_ {n=2\atop n\neq m+1}E_n\ketbra{n}{n}                                        \nonumber\\
     &=\frac{\hbar\omega_m}{2}\sigma^{(m)}_z+\sum^N_ {n=2\atop n\neq m+1}E_n\ketbra{n}{n}
\end{eqnarray}
and the interaction Hamilton is
\begin{eqnarray}\label{interact-hamilton-s1}
H^{(m)}=d_m(\ketbra{m+1}{1}+\ketbra{1}{m+1})
    =d_m\sigma^{(m)}_x.
\end{eqnarray}
So the total Hamilton becomes
\begin{eqnarray}
         H & = H_0+H^{(m)}   \nonumber\\
         &=\frac{\hbar\omega_m}{2}\sigma^{(m)}_z+\sum^N_ {n=2\atop n\neq m+1}E_n\ketbra{n}{n}
         +d_m\sigma^{(m)}_x   \nonumber\\
         & =H_c^{(m)}+\sum^N_ {n=2\atop n\neq m+1}E_n\ketbra{n}{n},
\end{eqnarray}
where $H_c^{(m)}=\frac{\hbar\omega_m}{2}\sigma^{(m)}_z+d_m\sigma^{(m)}_x$.

As $H$ does not depend on $t$,
the time evolution operator can be written as
\begin{eqnarray}\label{T-evolution-op}
 U^{(m)}(t)&=&e^{-iHt/\hbar}        \nonumber          \\
          &=&\exp\left[-\frac{it}{\hbar}\left(H_c^{(m)}+\sum^N_ {n=2\atop n\neq m+1}E_n\ketbra{n}{n}\right)\right] \nonumber      \\
         & =&\left[ \cos(\Omega_mt/\hbar)-\frac{i}{\Omega_m}\sin(\Omega_mt/\hbar)H_c^{(m)}\right ] \nonumber \\
         && \times \exp\left[-i\sum^N_ {n=2\atop n\neq m+1}E_n\ketbra{n}{n}t/\hbar\right],
\end{eqnarray}
from the fact                                                                                                     \begin{eqnarray}
  \left(H_c^{(m)}\right)^{2n}=\Omega_m^{2n}I ,   \qquad
  \left(H_c^{(m)}\right)^{2n+1}=\Omega_m^{2n}H_c^{(m)},
\end{eqnarray}
where $\Omega_m=\sqrt{\left(\frac{\hbar\omega_m}{2}\right)^2+d_m^2}$.

For convenience, it won't change the final results naturally, if we let $\hbar=1$. So
\begin{eqnarray}\label{simplified-T-op}
  U^{(m)}(t)   &=&  \left[ \cos(\Omega_mt)-\frac{i}{\Omega_m}\sin(\Omega_mt)H_c^{(m)}\right ] \nonumber \\
   && \times \exp\left[-i\sum^N_ {n=2\atop n\neq m+1}E_n\ketbra{n}{n}t\right], \ \ (m\geq 2)
\end{eqnarray}
further more, the evolution of system in each cycle can be investigated in schr\"{o}dinger picture, which is more convenient than the method using a changing field.

\subsection{Cycle 1}

In the first cycle, the initial state of the system  is $|1\rangle$.
Noticing that (\ref{simplified-T-op}) exclude the case $m=1$,  we has to calculate $U^{(1)}(t)$ in cycle-1, written as
\begin{eqnarray}\label{cycle-T-op}
   U^{(1)}(t) =\left[ \cos(\Omega_1t)-\frac{i}{\Omega_1}\sin(\Omega_1t)H_c^{(1)}\right ]
   \exp\left[-i\sum^N_ {n=3}E_n\ketbra{n}{n}t\right].
\end{eqnarray}
Then after time period $\tau_1$ in step one, system's state becomes
\begin{eqnarray}\label{cycle-1-field-sta}
  \ket{\psi_1} &=& U^{(1)}(\tau_1) \ket{1} \nonumber \\
  &=& \cos(\Omega_1\tau_1)\ket{1}+\frac{i\hbar\omega_1}{2\Omega_1}\sin(\Omega_1\tau_1)\ket{1}
  -\frac{i d_1}{\Omega_1}\sin(\Omega_1\tau_1)\ket{2}
\end{eqnarray}
when $d_1\gg \hbar\omega_1$, i.e. the magnitude of external field is very large. Then $\hbar\omega_1/\Omega_1\rightarrow 0,\; d_1/\Omega_1\rightarrow 1$, we get
\begin{equation}\label{state-of-c1}
 \ket{\psi_1}=\cos(\Omega_1\tau_1)\ket{1}-i\sin(\Omega_1\tau_1)\ket{2}.
\end{equation}
Therefore after free evolution for $\tau'_1$ time period, the final state is
\begin{equation}
 \ket{\psi'_1}=
 e^{-iH_0\tau'_1}\ket{\psi_1}=\coea{1}{1}\ket{1}+\coea{1}{2}\ket{2}.
\end{equation}
with
\begin{eqnarray}\label{c_1-coefficient}
   &\coea{1}{1}= \cos(\Omega_1\tau_1)e^{-iE_1\tau'_1} \\
   &\coea{1}{2}= -i\sin(\Omega_1\tau_1)e^{-iE_2\tau'_1}.
\end{eqnarray}

\subsection{Cycle 2}

For cycle-2, the initial state is the final state of cycle 1. Firstly,
we apply the control field for time period $\tau_2$. Using (\ref{simplified-T-op}) for $m=2$ and assuming $d_2\gg \hbar\omega_2$ is satisfied, we obtain the state
\begin{eqnarray}\label{cycle-2-f}
   \ket{\psi_2}&=&  U^{(2)}(\tau_2)\ket{\psi'_1}  \nonumber\\
    &=& \cos(\Omega_2\tau_2)\coea{1}{1}\ket{1}+\cos(\Omega_2\tau_2)e^{-iE_2\tau_2}\coea{1}{2}\ket{2} \nonumber \\
    & & -i\sin(\Omega_2\tau_2)\coea{1}{1}\ket{3}.
\end{eqnarray}
We turn off the external field, then after free system evolution for time period $\tau'_2$, the final state is
\begin{eqnarray}\label{c2-sta-final}
  \ket{\psi'_2}&= e^{-iH_0\tau'_2}\ket{\psi_2}
   =\coea{2}{1}\ket{1}+\coea{2}{2}\ket{2}+\coea{2}{3}\ket{3},
\end{eqnarray}
where
\begin{eqnarray}\label{c2-coeffict}
  & \coea{2}{1}=\cos(\Omega_2\tau_2)e^{-iE_1\tau'_2}\coea{1}{1} \nonumber\\
  & \coea{2}{2}=\cos(\Omega_2\tau_2)e^{-iE_2(\tau_2+\tau'_2)}\coea{1}{2}  \nonumber\\
  &\coea{2}{3}=-i\sin(\Omega_2\tau_2)e^{-iE_3\tau'_2}\coea{1}{1}
\end{eqnarray}

\subsection{From $(m-1)$-th cycle to $m$-th cycle}
To figure out the explicit expression of Target state when all the control processes end. The recursion relation between coefficients of the $(m-1)$-th cycle and $m$-th cycle is need. Suppose after $(m-1)$-th cycle, we write down the state as
\begin{equation}\label{m-1-cycle-sta}
  \ket{\psi'_{m-1}}=\sum_{k=1}^m \coea{m-1}{k}\ket{k}.
\end{equation}
Interacting with control field for time period $\tau_m$ and the restrictions $d_m\gg \hbar\omega_m$, which lead to $\hbar\omega_m/\Omega_m\rightarrow 0,\; d_m/\Omega_m\rightarrow 1$, we have the state
\begin{eqnarray}\label{m-cycle-s}
   \ket{\psi_m}&= \cos(\Omega_m\tau_m)\coea{m-1}{1}\ket{1} \nonumber \\
  &\,\quad+\sum^m_{k=2}\cos(\Omega_m\tau_m)e^{-iE_k\tau_m}\coea{m-1}{k}\ket{k}  \nonumber\\
  &\,\quad-i\sin(\Omega_m\tau_m)\coea{m-1}{1}\ket{m+1}.
\end{eqnarray}
After free evolution for time period $\tau'_m$, the final state of the $m$-th cycle is
\begin{eqnarray}\label{final-m-cycle-s}
  \ket{\psi'_m}   =e^{-iH_0\tau'_m}\ket{\psi_m}
    =\sum^{m+1}_{k=1}\coea{m}{k}\ket{k}.
\end{eqnarray}
where the coefficients are
\begin{eqnarray}
   & \coea{m}{1}=\cos(\Omega_m\tau_m)e^{-iE_1\tau'_m}\coea{m-1}{1},  \label{recursion-2way-1} \\
   & \coea{m}{k}=\cos(\Omega_m\tau_m)e^{-iE_k(\tau_m+\tau'_m)}\coea{m-1}{k},  \quad 2\leq k \leq m, \label{recursion-2way-2}    \\
   & \coea{m}{m+1}=-i\sin(\Omega_m\tau_m)e^{-iE_{m+1}\tau'_m}\coea{m-1}{1},    \label{recursion-2way-3}
\end{eqnarray}
hereafter. Eqs.(\ref{recursion-2way-1}-\ref{recursion-2way-3}) establishes the relationship between the probability
amplitudes of the $(m-1)$-th cycle and the $m$-th cycle.

\subsection{Target state}
The target state of 2 and 3 dimensional system has be given out in subsection B and C. The case $N\geq 4$ will be investigated in rest of this section. From \ref{c_1-coefficient} and \ref{recursion-2way-1}, we can find that
\begin{equation}\label{m-cycle-1-coef}
  \coea{m}{1}=\prod^m_{i=1}\cos(\Omega_i\tau_i)\exp\left[-iE_1\sum^m_{i=1}\tau'_i\right]
\end{equation}
As $\coea{m}{m+1}$ just depends on $\coea{m-1}{1}$, it can be written down as
\begin{eqnarray}\label{coeffit-m-cycle-st-m+1}
  \coea{m}{m+1}&=&-i\sin(\Omega_m\tau_m)\prod^{m-1}_{i=1}\cos(\Omega_i\tau_i) \nonumber \\
   &&\times \exp\left[-iE_1\sum^{m-1}_{i=1}\tau'_i+E_{m+1}\tau'_m\right].
\end{eqnarray}
For the coefficient $\coea{m}{2}$, using (\ref{c2-coeffict}), we obtain
\begin{eqnarray}\label{m-cycle-st-2-coef}
  \coea{m}{2} &=& -i\sin(\Omega_1\tau_1)\prod^m_{i=2}\cos(\Omega_i\tau_i) \nonumber \\
    && \times \exp\left\{-iE_2\left[\sum^m_{i=2}(\tau_i+\tau'_i)+\tau'_1\right]\right\}.
\end{eqnarray}
To derive coefficient $a^{(m)}_{k}$, $3\leq k \leq m$, we use
\begin{eqnarray}\label{coef-k-k-1}
   \coea{k-1}{k} &=& -i\sin(\Omega_{k-1}\tau_{k-1}) \nonumber \\
   && \times \prod^{k-2}_{i=1}\cos(\Omega_i\tau_i)
    \exp\left[-iE_1\sum^{k-2}_{i=1}\tau'_i+E_k\tau'_{k-1}\right],
\end{eqnarray}
which is obtained from (\ref{coeffit-m-cycle-st-m+1}) by replacing $m$ by $k-1$.
Then we can recursively have
\begin{eqnarray}\label{coef-m-cycle-k-sta}
\coea{m}{k}
&=\prod^m_{i=k}\cos(\Omega_i\tau_i)\exp\left[-iE_k\sum^{m}_{i=k}\tau_i+\tau'_i\right]\coea{k-1}{k} \nonumber \\
&=-i\sin(\Omega_{k-1}\tau_{k-1})\prod^m_{i=1\atop i\neq k+1}\cos(\Omega_i\tau_i) \nonumber\\
&\,\quad \times \exp\left[-iE_1\sum^{k-2}_{i=1}\tau'_i+E_k\left(\sum^{m}_{i=k}\tau_i+\sum^{m}_{i=k-1}\tau'_i\right)\right].
\end{eqnarray}
From expression (\ref{coef-m-cycle-k-sta}), Notice that it does not recover the
case $k=2$, only to be valid for $k=m+1$. So far, all the probability amplitudes
after $m$-th cycle are given by \ref{m-cycle-1-coef}, \ref{m-cycle-st-2-coef}
and \ref{coef-m-cycle-k-sta}

Since the system we considered here is $N$ dimension, we implement $N-1$ cycles to
drive the system to arbitrary target states. Thus letting $m=N-1$, we obtain the probability
amplitude of the target state
\begin{eqnarray}\label{N-1-cycle-coef}
     \coea{N-1}{1}&=&\prod^{N-1}_{i=1}\cos(\Omega_i\tau_i)\exp\left[-iE_1\sum^{N-1}_{i=1}\tau'_i\right],   \nonumber     \\
     \coea{N-1}{2}& =&-i\sin(\Omega_1\tau_1)\prod^{N-1}_{i=2}\cos(\Omega_i\tau_i)
                     \exp\left\{-iE_2\left[\sum^{N-1}_{i=2}T_i+\tau'_1\right]\right\}, \nonumber\\
     \coea{N-1}{k}&=&-i\sin(\Omega_{k-1}\tau_{k-1})\prod^{N-1}_{i=1\atop i\neq k+1}\cos(\Omega_i\tau_i) \nonumber\\
     && \times
     \exp\left[-iE_1\sum^{k-2}_{i=1}\tau'_i+E_k\left(\sum^{N-1}_{i=k}\tau_i+\sum^{N-1}_{i=k-1}\tau'_i\right)\right],\nonumber\\
     && \quad (3\leq k\leq N).
\end{eqnarray}

\subsection{Control parameters}

For a control problem, the target state, or in other words, the
amplitude $a^{N-1}_m$ of the target state, is given. What we need to do is to
determine the control parameters
$\{ \tau_i, \tau'_i \ |\ i=1,2,...,N-1 \}$ from the probability amplitude
of the target state. For convenience, we write the target state as
\begin{equation}\label{last-state-for-short}
     \ket{\psi}=\sum^N_{n=1}\gamma_n C_n \ket{n}
\end{equation}
where $C_n$s are the real part of the amplitude
\begin{eqnarray}
     C_1&=\prod^{N-1}_{i=1}\cos(\Omega_i\tau_i), \label{coefficient-Nc-1}\\
     C_2&=\sin(\Omega_1\tau_1)\prod^{N-1}_{i=2}\cos(\Omega_i\tau_i),  \label{coefficient-Nc-2} \\
     C_n&=\sin(\Omega_{k-1}\tau_{k-1})\prod^{N-1}_{i=1\atop i\neq k+1}\cos(\Omega_i\tau_i) ,\! 3\leq n\leq N,
      \label{coefficient-Nc-3}
\end{eqnarray}
and $\gamma_n$s are phases
\begin{eqnarray}\label{coefficient-phase}
   \gamma_1 = & \exp\left[-iE_1\sum^{N-1}_{i=1}\tau'_i\right] \nonumber\\
   \gamma_2 = & \exp\left\{-iE_2\left[\sum^{N-1}_{i=2}(\tau_i+\tau'_i)+\tau'_1\right]\right\}\nonumber\\
   \gamma_n = & \exp\left[-iE_1\sum^{k-2}_{i=1}\tau'_i+E_k\left(\sum^{N-1}_{i=k}\tau_i+\sum^{N-1}_{i=k-1}\tau'_i\right)-\frac{\pi i}{2}\right] \nonumber\\
    & (3\leq n\leq N).
\end{eqnarray}

For a given target state, equivalently, $C_n$ and $\gamma_n$ are given,
we can calculate control parameters $\{\tau_n,\tau'_n|n=1,2,...,N-1  \}$.
From (\ref{coefficient-Nc-1}),(\ref{coefficient-Nc-2}) and $C_1,C_2$, we can determine $\tau_1$.
Then form (\ref{coefficient-Nc-3}) with $n=3$ ,(\ref{coefficient-Nc-1}) and $C_1,C_3$, we can obtain
$\tau_2$. Repeating this process, we can find out
all parameters $\tau_n$, $n=1,2,...,N-1$, determining the coupling time between the system and field.

All $\tau^\prime_i$ can be obtained from (\ref{coefficient-phase}).
From $\gamma_1$ and $\gamma_2$, we can obtain
\begin{equation}
  E_1\sum_{i=1}^{N-1} \tau'_i, \qquad E_2 \sum_{i=2}^{N-1}\tau_i + E_2 \tau'_1,
\label{coefficient-003}
\end{equation}
from which we get the value of $\tau'_1$ and $\sum_{i=2}^{N-1}\tau'_i$. From $\gamma_4$, we
find
\begin{equation}
E_1\sum^2_{i=1}\tau'_i+E_4\left(\sum^{N-1}_{i=4}\tau_i+\sum^{N-1}_{i=3}\tau'_i\right),
\end{equation}
from which as well as  $\tau'_1$ and $\sum_{i=2}^{N-1}\tau'_i$, we can obtain $\tau'_2$.
Repeating this process, we can obtain all $\tau'_i$ i.e. all the time for free evolution.

\section{System II}
\subsection{Time evolution operator}

For system II, the interaction operator is given in
E.q.(\ref{interaction-opterator-s2}). This operator
is same as that for system I except the state $\ket{1}$ is
replaced by $\ket{m}$. So we can follow exactly the same
procedure as in last section, namely, to simplify the time evolution operator

 In the $m-th$ cycle, We define
\begin{equation}
   \hbar\tomega_m=E_{m+1}-E_m
\end{equation}
and reset the energy base point $\frac{1}{2}(E_{m+1}+E_m)=0$.
Using these two conditions in each cycle, we get the simplified free Hamilton:
\begin{eqnarray}\label{h0-hamilton-s2}
  H_0 &=& \sum_n^N E_n\ketbra{n}{n}                                                       \nonumber\\
     &=&E_m\ketbra{m}{m}+E_ {m+1}\ketbra{m+1}{m+1}                                        \nonumber  \\
     & &+\sum^N_ {n=1\atop n\neq m,m+1}E_n\ketbra{n}{n}                                      \nonumber  \\
     &=& -\frac{\hbar\tomega_m}{2}\ketbra{m}{m}+\frac{\hbar\tomega_m}{2}\ketbra{m+1}{m+1}        \nonumber\\
     & &+\sum^N_ {n=1\atop n\neq m,m+1}E_n\ketbra{n}{n}                                        \nonumber\\
     &=&\frac{\hbar\tomega_m}{2}\tsigma^{(m)}_z+\sum^N_ {n=2\atop n\neq m+1}E_n\ketbra{n}{n}
\end{eqnarray}
and the interaction Hamilton is
\begin{eqnarray}\label{interact-hamilton-s2}
H^{(m)}=d_m(\ketbra{m+1}{m}+\ketbra{m}{m+1}) 
    =d_m\tsigma^{(m)}_x.
\end{eqnarray}
Thus the total Hamilton takes the form:
\begin{eqnarray}
         H & = \frac{\hbar\tomega_m}{2}\tsigma^{(m)}_z+\sum^N_ {n=1\atop n\neq m,m+1}E_n\ketbra{n}{n}
               +d_m\tsigma^{(m)}_x     \nonumber \\
         & =H_c^{'(m)}+\sum^N_ {n=1\atop n\neq m,m+1}E_n\ketbra{n}{n},
\end{eqnarray}
where 
\begin{equation}
H_c^{'(m)}=\frac{\hbar\tomega_m}{2}\tsigma^{(m)}_z+d_m\tsigma^{(m)}_x.
\end{equation}

With respect to the fact that the control field is constant, i.e. $H$ dosen't contain $t$, 
the time evolution operator can be obtained as
\begin{eqnarray}\label{T-evolution-op-s2}
 U^{(m)}(t)
          &=&\exp\left[-i\left(H_c^{'(m)}+\sum^N_ {n=1\atop n\neq m,m+1}E_n\ketbra{n}{n}\right)t\right] \nonumber      \\
          &=&\left[ \cos(\Omega_mt)-\frac{i}{\Omega_m}\sin(\Omega_mt)H_c^{'(m)}\right ]   \nonumber     \\
          && \times\exp\left[-i\sum^N_ {n=1\atop n\neq m,m+1}E_n\ketbra{n}{n}t\right],\quad m\geq 2
\end{eqnarray}
using the fact                                                                                                     \begin{eqnarray}
  &\left(H_c^{'(m)}\right)^{2n}=\Omega_m^{2n}I,    \nonumber \\
  &\left(H_c^{'(m)}\right)^{2n+1}=\Omega_m^{2n}H_c^{'(m)}.
\end{eqnarray}

\subsection{Determine amplitude $a_m$}

From (\ref{T-evolution-op-s2}) for the case $m=1$, $U^{(1)}(t)$ is written as
\begin{eqnarray}\label{cycle-T-op}
   U^{(1)}(t) &=& \left[ \cos(\Omega_1t)-\frac{i}{\Omega_1}\sin(\Omega_1t)H_c^{'(1)}\right ]  \nonumber \\
   && \times \exp\left[-i\sum^N_ {n=3}E_n\ketbra{n}{n}t\right].
\end{eqnarray}
For this model, cycle 1 is exactly the same as the system I.
So after the cycle 1, the system is driven to the state
\begin{eqnarray}\label{final-sta-c1-s2}
\ket{\psi'_1}=\coea{1}{1}\ket{1}+\coea{1}{2}\ket{2},
\end{eqnarray}
where
\begin{eqnarray}\label{initiala-Condition-s2}
   \coea{1}{1}=&\cos(\Omega_1\tau_1)e^{-iE_1\tau'_1},\nonumber \\
   \coea{1}{2}=&-i\sin(\Omega_1\tau_1)e^{-iE_2\tau'_1}.
\end{eqnarray}

Different from system I, in cycle 2, the control field $f(t)=d_2$ causes transition
between $\ket{2}$ and $\ket{3}$. We can find the state after the cycle 2 in
Schr\"{o}dinger picture as
\begin{equation}\label{short-Form-c2-s2}
  \ket{\psi'_2}=
  \coea{2}{1}\ket{1}+\coea{2}{2}\ket{2}+\coea{2}{3}\ket{3},
\end{equation}
where
\begin{eqnarray}\label{coefficient-c2-s2}
  \coea{2}{1}=&\cos(\Omega_2\tau_2)e^{-iE_1(\tau_1+\tau'_1)}\coea{1}{1},  \\
  \coea{2}{2}=&\cos(\Omega_2\tau_2)e^{-iE_2 \tau'_2}\coea{1}{2}, \\
 \coea{2}{3}=&-i\sin(\Omega_2\tau_2)e^{-iE_3 \tau'_2}\coea{1}{2}.
\end{eqnarray}

To obtain the target state, we first deduce the recursion relations
between the $(m-1)$-th cycle and the $m$-th cycle. To this end,
we suppose that, after $m-1$ cycles, the system is on the state
\begin{equation}\label{State-cm-s2}
  \ket{\psi'_{m-1}}= \sum_{k=1}^m \coea{m-1}{k}\ket{k}.
\end{equation}
Then after interactions with the control field for
time period $\tau_m$, and free evolution for time period $\tau'_m$,
we find the final state after cycle $m$ as
\begin{equation}\label{state-cm-s2}
  \ket{\psi'_m}=\sum_{k=1}^{m+1} \coea{m}{k}\ket{k},
\end{equation}
with ($m \geq 2$)
\begin{eqnarray}
  &&\coea{m}{k}=\cos(\Omega_m\tau_m)e^{-iE_k(\tau_m+\tau'_m)}\coea{m-1}{k},\quad 1\leq k\leq m-1,
               \label{induction-S2-1}\\
  &&\coea{m}{m}= \cos(\Omega_m\tau_m)e^{-iE_m \tau'_m} \coea{m-1}{m},  \\
  &&\coea{m}{m+1}=-i\sin(\Omega_m\tau_m)e^{-iE_{m+1}\tau'_m}\coea{m-1}{m}.\label{all-Coeff-cm-s2}
\end{eqnarray}
From those recursion relations, and initial conditions (\ref{initiala-Condition-s2})
we can get all the explicit expressions of $\coea{m}{k}$.
It is easy to see that
\begin{eqnarray}\label{all-Coef-cm-1-2-s2}
  \coea{m}{1}&=& \prod^m_{i=1}\cos(\Omega_i\tau_i)\exp\left\{-iE_1\left[\sum^m_{i=2}(\tau_i+\tau'_i)+\tau'_1 \right]\right\},  \\
  \coea{m}{2}&=&-i\sin(\Omega_1\tau_1)\prod^m_{i=2}\cos(\Omega_i\tau_i) \nonumber  \\
  && \times\exp\left\{-iE_2\left[\sum^m_{i=3}(\tau_i+\tau'_i)+\tau'_2+\tau'_1 \right]\right\},   \\
  \coea{m}{m+1}&=&-i\sin(\Omega_m\tau_m)e^{-iE_{m+1}\tau'_m} \coea{m-1}{m}    \nonumber    \\
  &= &  \cdots \nonumber \\
  &= & \left(-i\right)^{m-1}\prod^m_{i=2}\sin(\Omega_i\tau_i)
      \times\exp\left[-i\left(
      \sum^m_{i=2}E_{i+1}\tau'_i\right)\right]\coea{1}{1} \nonumber\\
  &=&\left(-i\right)^m\prod^m_{i=1}\sin(\Omega_i\tau_i) \exp\left[-i\left(
      \sum^m_{i=1}E_{i+1}\tau'_i\right)\right].  
  \label{coef-m-m+1-s2}
\end{eqnarray}
Then using \ref{induction-S2-1} and $\coea{m-1}{m}$, which is obtained
from (\ref{coef-m-m+1-s2}) by replacing $m$ by $m-1$, we have
\begin{eqnarray}\label{generalForm-cm-staK-s2}
  \coea{m}{k}
   &= & \prod^m_{i=k+1}\cos(\Omega_i\tau_i)\exp\left[-iE_k\sum^m_{i=k+1}(\tau_i+\tau'_i)\right]\nonumber\\
   && \cos(\Omega_k\tau_k)e^{-iE_k\tau'_k}\coea{k-1}{k}  \nonumber \\
   &=& \prod^m_{i=k}\cos(\Omega_i\tau_i)\exp\left\{-iE_k\left[\sum^m_{i= k+1}(\tau_i+\tau'_i)+\tau'_k\right]\right\}\nonumber \\
   &&\times \left(-i\right)^{k-1}\prod^{k-1}_{i=1}\sin(\Omega_i\tau_i) \exp\left[-i\left(\sum^{k-1}_{i=1}E_{i+1}\tau'_i\right)\right]\nonumber\\
   &=&\left(-i\right)^{k-1}\prod^m_{i=k}\cos(\Omega_i\tau_i)\prod^{k-1}_{i=1}\sin(\Omega_i\tau_i)\nonumber \\
   && \times\exp\left\{-i E_k\left[\sum^m_{i= k+1}(\tau_i+\tau'_i)+\tau'_k\right]-i\sum^{k-1}_{i=1}E_{i+1}\tau'_i \right\}.
\end{eqnarray}
One can check that (\ref{generalForm-cm-staK-s2}) includes the case $k=2$ and $k=m$
as special cases.

Therefore, after $N-1$ cycles, we arrive at the target state
\begin{eqnarray}\label{finalFormOfcoefficient}
    \coea{N-1}{1}&=&\prod^{N-1}_{i=1}\cos(\Omega_i\tau_i)\exp\left\{-iE_1\left[\sum^{N-1}_{i=2}(\tau_i+\tau'_i)+\tau'_1 \right]\right\}, \nonumber \\
    \coea{N-1}{n}&=&\left(-i\right)^{n-1}\prod^{N-1}_{i=n}\cos(\Omega_i\tau_i)\prod^{n-1}_{i=1}\sin(\Omega_i\tau_i) \nonumber \\
    &&\times\exp \left\{-i E_n\left[\sum^{N-1}_{i=n+1}(\tau_i+\tau'_i)+\tau'_n\right]-i\sum^{n-1}_{i=1}E_{i+1}\tau'_i \right\},\\
    && \quad \quad\quad \quad (2\leq m\leq N-1) ,\nonumber \\
 \coea{N-1}{N}&=&\left(-i\right)^{N-1}\prod^{N-1}_{i=1}\sin(\Omega_i\tau_i) \exp\left[-i\left(\sum^{N-1}_{i=1}E_{i+1}\tau'_i\right)\right].
\end{eqnarray}

\subsection{Control parameters}

To determine control parameters $\tau_i,\tau'_i, 1\leq i\leq N-1$, we write
the target state as
\begin{equation}\label{stateSuperposition}
     \ket{\psi}=\sum^N_{n=1}a_n \ket{n}=\sum^N_{n=1}\gamma_n C_n \ket{n},
\end{equation}
in which
\begin{eqnarray}\label{realProbabilityAmplitude}
     C_1&=\prod^{N-1}_{i=1}\cos(\Omega_i\tau_i),  \nonumber \\
     C_m&=\prod^{N-1}_{i=m}\cos(\Omega_i\tau_i)\prod^{m-1}_{i=1}\sin(\Omega_i\tau_i) ,\quad (2\leq m\leq N-1), \nonumber\\
     C_N&=\prod^{N-1}_{i=1}\sin(\Omega_i\tau_i),
\end{eqnarray}
and phase $\gamma_n$
\begin{eqnarray}\label{relativePhase}
   \gamma_1&= \exp\left\{-iE_1\left[\sum^{N-1}_{i=2}(\tau_i+\tau'_i)+\tau'_1 \right]\right\} ,  \nonumber \\
   \gamma_m &= \left(-i\right)^{n-1}
     \exp \left\{-i E_m\left[\sum^{N-1}_{i=\atop m+1}(\tau_i+\tau'_i)+\tau'_m\right]-i\sum^{m-1}_{i=1}E_{i+1}\tau'_i \right\},  \nonumber  \\
   \gamma_N&= \left(-i\right)^{N-1}\exp\left[-i\left(\sum^{N-1}_{i=1}E_{i+1}\tau'_i\right)\right].
\end{eqnarray}

For a given target state, namely, $C_n$ and $\gamma_n$ are given, we can determine
the control parameters $\{\tau_n,\tau'_n|n=1,2,...,N-1  \}$. From $C_1,C_2$ we can determine
$\tau_1$, and then from $C_2,C_3$ we obtained $\tau_2$. Recursively we can
obtain all $\tau_n$.

For $\tau'_n$, from $\gamma_1$ and $\gamma_2$, we obtain
\begin{eqnarray}
E_1\sum_{i=2}^{N-1} \tau'_i+E_1\tau'_1, \qquad E_2\sum_{i=2}^{N-1}\tau'_i+E_2 \tau'_1,
\end{eqnarray}
respectively. As $E_1\neq E_2$, we obtain $\tau'_1$ and $\sum_{i=2}^{N-1}\tau'_i$.
From $\gamma_3,\gamma_4$, we can obtain
\begin{equation}
E_3\sum_{i=2}^{N-1} \tau'_i+E_2\tau'_1, \qquad E_4\sum_{i=3}^{N-1}\tau'_i+\sum_{i=1}^2 E_{i+1} \tau'_i,
\end{equation}
from which we obtain $\tau'_2$ as well as $\sum_{i=3}^{N-1}\tau'_i$. Repeating this
process, we can obtain all $\tau'_i$ .

\section{Conclusion}

In this paper we proposed a protocol to drive two types of  finite
dimensional quantum system to an arbitrary given target states. The control
parameters are time periods $\{\tau_m, \tau^\prime_m | m=1,2,\cdots,N-1\}$
which can be explicitly determined from the probability
amplitudes of the given target states. Relationship between control parameters
and amplitudes is trigonometric  functions and can be solved explicitly.

We have $2(N-1)$ real control parameters. In the target state there are $N$
complex or $2N$ real parameters. Taking into account the normalization
condition of target state, one has $2(N-1)$ real parameters, the same as
the number of the real control parameters. From this fact we can
conclude that we can drive the system to an arbitrary target state by
choosing appropriate control parameters $\{\tau_m, \tau^\prime_m\}$.

As further works, we would like to consider the indirect control protocol of
finite quantum system by generalizing the control scheme in this paper. We
also would like to consider the control protocol in the presentence of
environment.

\section*{Ackonwledgement}

This work is supported by the National Science Foundation of China under grand
number 11075108.

\end{document}